\documentclass[twocolumn]{aastex63}

\usepackage[caption=false]{subfig}
\usepackage{mathtools,amssymb}
\newcommand{\xiinj}{\xi_{\rm inj}}
\newcommand{\vsh}{v_{\rm sh}}
\newcommand{\nism}{n_{\rm ISM}}
\newcommand{\mpr}{m_{\rm p}}

\newcommand{\rt}{R_{\rm tot}}

\newcommand{\rs}{R_{\rm sub}}
\newcommand{\pinj}{p_{\rm inj}}
\newcommand{\xicr}{\xi_{\rm CR}}
\newcommand{\xib}{\xi_{\rm B}}

\newcommand{\w}[1]{v_{\rm A,#1}}

\submitjournal{ApJ}

\shorttitle{Steep Spectra with Revised DSA}

\begin{document}

\title{Steep Cosmic Ray Spectra with Revised Diffusive Shock Acceleration}

\correspondingauthor{Rebecca Diesing}
\email{rrdiesing@uchicago.edu}

\author[0000-0002-6679-0012]{Rebecca Diesing}
\affiliation{Department of Astronomy and Astrophysics, The University of Chicago, 5640 S Ellis Ave, Chicago, IL 60637, USA}

\author[0000-0003-0939-8775]{Damiano Caprioli}
\affiliation{Department of Astronomy and Astrophysics, The University of Chicago, 5640 S Ellis Ave, Chicago, IL 60637, USA}
\affiliation{Enrico Fermi Institute, The University of Chicago, Chicago, IL 60637, USA}

\begin{abstract}

Galactic cosmic rays (CRs) are accelerated at the forward shocks of supernova remnants (SNRs) via diffusive shock acceleration (DSA), an efficient acceleration mechanism that predicts power-law energy distributions of CRs. 
However, observations of nonthermal SNR emission imply CR energy distributions that are generally steeper than $E^{-2}$, the standard DSA prediction. 
Recent results from kinetic hybrid simulations suggest that such steep spectra may arise from the drift of magnetic structures with respect to the thermal plasma downstream of the shock. 
Using a semi-analytic model of non-linear DSA, we investigate the  implications that these results have on the phenomenology of a wide range of SNRs. 
By accounting for the motion of magnetic structures in the downstream, we produce CR energy distributions that are substantially steeper than $E^{-2}$ and consistent with observations. Our formalism reproduces both modestly steep spectra of Galactic supernova remnants ($\propto E^{-2.2}$) and the very steep spectra of young radio supernovae ($\propto E^{-3}$). \\

\end{abstract}

\section{Introduction} \label{sec:intro}

Understanding the origin of Galactic cosmic rays (CRs) with energies up to $\sim 10^8$ GeV requires a complete paradigm for their acceleration and propagation. The best source candidates for such acceleration are supernova remnants (SNRs), which provide sufficient energy and an efficient acceleration mechanism \citep{hillas05,berezhko+07,ptuskin+10,caprioli+10a}. 
In this mechanism, known as \emph{diffusive shock acceleration} (DSA), particles are scattered by magnetic field perturbations, resulting in diffusion across the SNR forward shock and an energy gain with each crossing \citep[]{fermi54, krymskii77, axford+77p, bell78a, blandford+78}. 

DSA predicts a power law momentum distribution of particles, $f_{\rm sh}(p) \propto p^{-q_{\rm p}}$, where $f_{\rm sh}(p)$ is the instantaneous momentum distribution of particles at the shock and $q_{\rm p}$ is set by the balance between the energy gained with each crossing and the escape of particles from the acceleration region \citep{bell78a}. 
Both of these quantities depend on the shock hydrodynamics such that $q_{\rm p}$ can be written in terms of the fluid compression ratio, $R = \rho_2/\rho_0$. Here, $\rho_1$ and $\rho_2$ are the densities of the fluid in front of the shock (upstream) and behind the shock (downstream) respectively. The relationship between $q_{\rm p}$ and $R$ reads,

\begin{equation}
    q_{\rm p} = \frac{3R}{R-1}.
\end{equation}
For a strong shock with Mach number $M \gg 1$, $R=4$ and we obtain $q_{\rm p} = 4$. Equivalently, DSA predicts power-law distributions in energy for relativistic particles, $\Phi_{\rm sh}(E) \propto E^{-q}$, where $\Phi_{\rm sh}(E)$ is the instantaneous energy distribution of particles at the shock. The relationship between $q$ and $R$ reads,

\begin{equation}
    q = \frac{R+2}{R-1},
\end{equation}
with $q = 2$ for a strong shock ($R=4$).

A modification to the standard DSA prediction arises when CRs carry a non-negligible fraction of the shock's energy. When this occurs, CRs can no longer be treated as test-particles, resulting in modifications to the shock hydrodynamics and thus particle spectra \citep[e.g., ][]{drury-volk81a, drury83, blandford+87,jones+91,berezhko+97,malkov+01,kang+05,kang+06,ellison+00,ellison+96,berezhko+99,amato+05,amato+06,caprioli+09a,caprioli+08}. 
In this non-linear DSA (NLDSA), the CR pressure produces a region in front of the shock where the fluid is compressed, heated, and slowed. The presence of this region, or \emph{precursor}, reduces the compression ratio near the shock into a subshock with $R_{\rm sub} \equiv \rho_2/\rho_1 < 4$. Meanwhile, the total compression ratio between the downstream and far upstream becomes larger than the standard prediction: $R_{\rm tot} \equiv \rho_2/\rho_0 > 4$. Note that, throughout this paper, subscripts 0, 1, 2, and 3 are used to denote quantities at upstream infinity, immediately upstream of the shock, immediately downstream of the shock, and far downstream respectively (see Figure \ref{fig:shock}).

As a result of these two compression ratios, NLDSA predicts concave CR spectra. More specifically, particles with lower energies remain close to the shock and probe $R_{\rm sub} < 4$, while particles with higher energies diffuse further upstream and probe $R_{\rm tot} > 4$. Thus, low/high energy particles are expected to exhibit spectra steeper/flatter than $E^{-2}$. The transition between these regimes occurs at the lowest energy where CRs carry non-negligible pressure, which is usually trans-relativistic. Since the nonthermal emission in astrophysical environments is typically generated by relativistic CRs, the classical NLDSA theory predicts that observations of non-thermal emission from shock-powered sources should be explained by CR spectra flatter than $E^{-2}$. 

\subsection{Theory vs. Observations}

This prediction is readily testable via observations of the nonthermal emission, e.g., from the relics of stellar explosions. 
However, the first GeV observations of SNRs, combined with preexisting TeV data, did not confirm the existence of concave spectra. 
On the contrary, they pointed toward CR acceleration with spectra \emph{steeper} than $E^{-2}$ \citep{caprioli11}. 
Notable examples include historical remnants such as Tycho's SNR \citep[$q = 2.3 \pm 0.2$, ][]{giordano+12,archambault+17} and Cassiopeia A \citep[$q = 2.36 \pm 0.02$ above 17 GeV, ][]{saha+14}. 

Further evidence for steep spectra comes in the form of SNR radio emission, particularly that of young, extragalactic supernovae (\emph{radio SNe}). These remnants exhibit synchrotron spectra that imply electron distributions as steep as $E^{-3}$ \cite[e.g., ][]{chevalier+06, chevalier+17, soderberg+10, soderberg+12, kamble+16}. However, it is possible that these synchrotron spectra probe the steep portion of a concave spectrum, since the electrons responsible are likely sub-GeV \citep[e.g., ][]{ellison+91, ellison+00, tatischeff09}.

The CR spectrum measured at Earth also points toward CR acceleration with spectra steeper than $E^{-2}$. In the standard picture of CR transport, this measured CR spectrum goes as $E^{-(q+\delta)}$, where $\delta$ is the slope of the CR residence time in the Galaxy: $\tau_{\rm res} \propto E^{-\delta}$. Measurements of the CR anisotropy suggest $\delta \sim 0.3$ \citep{blasi+11a, blasi+11b}. Meanwhile, secondary to primary ratios suggest that $0.3 \lesssim \delta \lesssim 0.4$, depending on the CR energy \citep[e.g., ][]{ams18}. Thus, fitting the observed Galactic CR spectrum--which goes as $E^{-2.7}$--requires $2.3 \lesssim q \lesssim 2.4$ \citep{evoli+19a, evoli+19b}.

\subsection{A Revised Theory of DSA}

A number of explanations for steep CR spectra have been proposed in the literature. These explanations include,

\begin{enumerate}
    \item Anisotropic or inhomogeneous CR transport \citep[e.g., ][and references therein]{kirk+96, bell+11} at fast oblique shocks ($\vsh \gtrsim 10^4$ km s$^{-1}$). While such transport could explain the steep spectra of radio SNe, it does not apply to quasi-parallel and/or slower shocks (i.e., older SNRs).
    
    \item Modified shock dynamics due to the presence of neutral hydrogen \citep[][]{blasi+12a, morlino+12b, morlino+13}  and/or steepening due to ion-neutral damping \citep{malkov+12}. While this idea may be consistent with the steep spectra of some SNRs propagating into a partially-ionized medium \cite[e.g.,][]{morlino+16}, it cannot explain radio SNe, since for fast shocks ($\vsh \gtrsim 3000$ km s$^{-1}$), ionization becomes dominant thereby eliminating the neutral return flux. 
    Moreover, it is unclear whether ion-neutral damping would produce a steepening or a low-energy cutoff around a few GeV that would effectively halt the acceleration process. 
    
    \item Effects arising from the convolution of multiple CR distributions, either over time \citep{malkov+19}, or space, specifically the convolution of spectra from regions in which the large scale magnetic field is either quasi-parallel or quasi-perpendicular to the shock normal \citep{hanusch+19}. However, both effects may only work with the inclusion of ad-hoc shock obliquities.
    Moreover, the former effect is predicated on a growing region in which the shock is quasi-parallel, and therefore cannot explain SNRs with magnetic field coherence lengths that are smaller than the size of the system (e.g., Tycho). 
    The latter would not apply in SNRs that probe a uniform background magnetic field (e.g., SN 1006). 
    
    \item CR energy loss in the upstream due to the generation of magnetic turbulence \citep{bell+19}. While such an effect may in principle steepen DSA spectra, it is not observed in kinetic simulations, as outlined in more detail below.
\end{enumerate} 
A more detailed summary of these explanations and their limitations can be found in \cite{caprioli+20}. 

Another possible explanation considers the role of the magnetic fluctuations responsible for CR scattering \cite[e.g.,][]{zirakashvili+08b,caprioli11,caprioli12,kang+18}. 
In the standard DSA theory, particles are isotropized in both the upstream and downstream such that they ``feel" a head-on collision with each crossing of the shock. The resulting energy gain per crossing thus depends on the difference in velocity between the upstream and downstream plasma, $u_1-u_2$. 
In reality, however, magnetic fluctuations--not thermal plasma--are responsible for particle scattering, meaning that particles will be isotropized in the \emph{fluctuation} frame. The relative drift between the fluid and the fluctuations was already present in the early DSA theory \citep{bell78a}, but it has been usually neglected because the fluctuation drift is roughly the Alfv\'en speed, much smaller than the fluid speed in the shock frame.
In the presence of CR-driven magnetic field amplification, however, such a drift may be significantly enhanced;
one can argue that, in the upstream, these fluctuations move against the fluid with the local Alfv\'en velocity in the amplified field, $\w1$ \citep[e.g.,][]{caprioli12}. 
Thus, CRs experience a smaller energy gain per crossing $\propto u_1-\w1-u_2$ or, equivalently, they ``feel" a compression ratio, $\tilde{R}$ that is smaller than that of the fluid,
\begin{equation}
    \tilde{R} = \frac{u_1-\w1}{u_2} < R = \frac{u_1}{u_2}.
\end{equation} 
This prescription may naturally lead to spectra that are steeper than $E^{-2}$ \citep{caprioli11,caprioli12}, and it has been used, e.g., to model for the broadband emission of Tycho's SNR \citep{morlino+12,slane+14} and of intracluster shocks \citep{kang+13}.

The potential role of such drifts had not been validated by self-consistent kinetic simulations until very recently, when \cite{haggerty+20} and \cite{caprioli+20} put forward unprecedentedly-long hybrid simulations (i.e., particle-in-cell simulations with kinetic ions and fluid electrons) that showed the onset of CR-modified shocks. That being said, the presence of a precursor is insufficient to explain the very steep spectra ($\propto E^{-3}$) of radio SNe, and its effect may be limited if magnetic field amplification in the upstream is spatially-dependent. 
In particular, if the local Alfv\'en speed decreases in the precursor, particles with long diffusion lengths will probe a region with reduced fluctuation drift, resulting in a flattening of the CR spectrum at the highest energies.

\begin{figure}
    \centering
    \includegraphics[width=0.5\textwidth, clip=true,trim= 50 40 25 50]{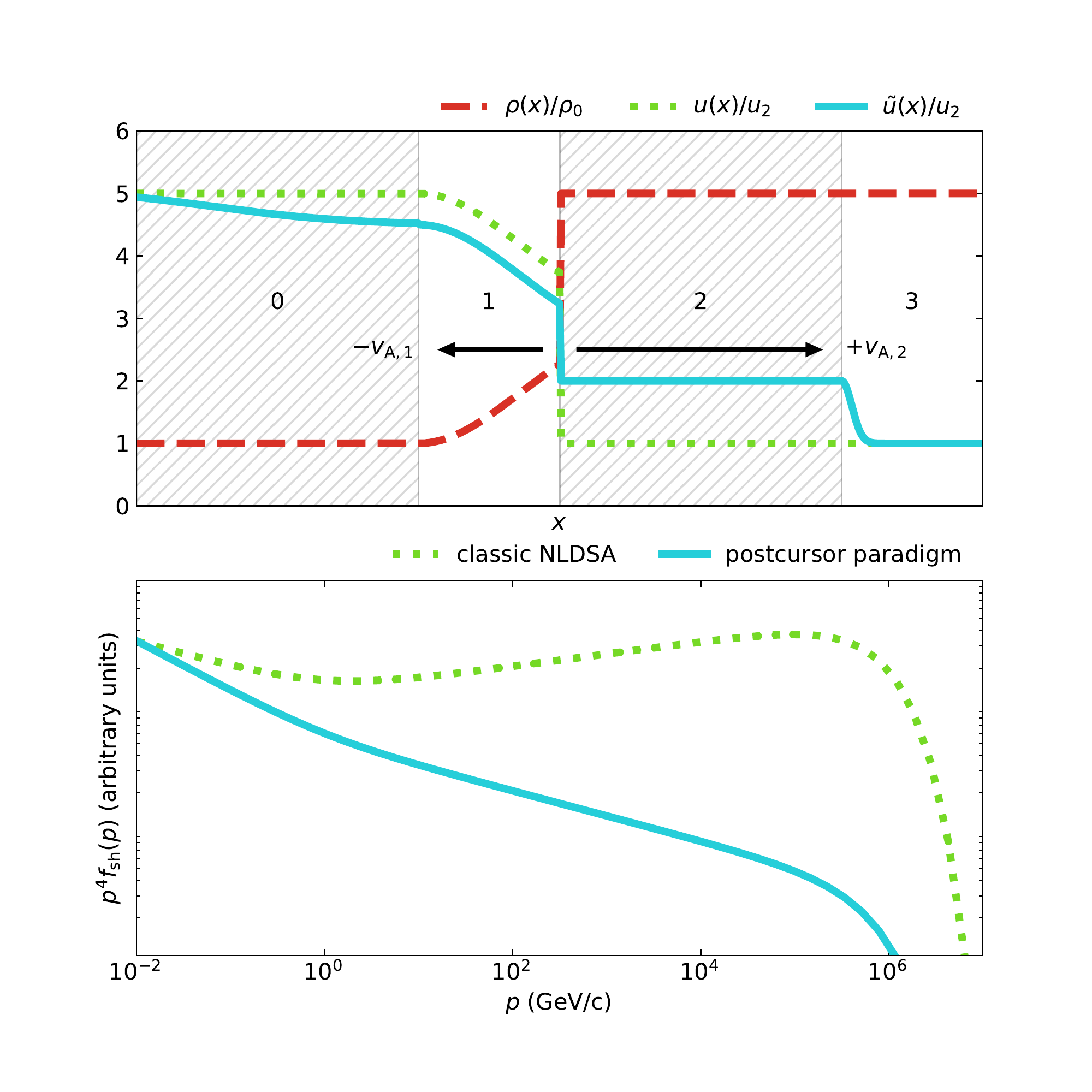}
    \caption{\emph{Top}: A sketch of the fluid density ($\rho(x)$; red dashed line), fluid velocity ($u(x)$; green dotted line), and the velocity of magnetic fluctuations ($\tilde{u}(x) = u(x)+v_{\rm A}(x)$; blue solid line) for a CR modified shock with precursor (Region 1) and postcursor (Region 2). Velocities are displayed in the shock rest frame. \emph{Bottom}: A sketch of the expected instantaneous particle distribution, $f_{\rm sh}(p)$, in the case with no net drift of magnetic fluctuations (classic NLDSA; green dotted line) and in the case with the net drifts shown in the top figure (postcursor paradigm; blue solid line). In the postcursor paradigm, particles experience an effective compression ratio smaller than that of the fluid, resulting in steeper spectra.}
    \label{fig:shock}
\end{figure}

However, \cite{haggerty+20} finds that not only does a precursor form in front of the shock, in which self-generated fluctuations move at roughly $\w1$ in the amplified field, but also that the motion of magnetic structures behind the shock leads to the formation of a \emph{postcursor}. 
In this picture, CR-driven magnetic fluctuations generated in the upstream retain their inertia over a non-negligible distance (larger than the CR diffusion length) when advected and compressed into the downstream. 
As a result, these fluctuations move away from the shock faster than the background plasma, or more specifically, with velocity $\tilde{u}_2 = u_2 + \w2$ with respect to the shock. 
A sketch of a CR-modified shock, with precursor and postcursor included, is shown in Figure \ref{fig:shock}.

Since CRs tend to isotropize with magnetic fluctuations, they too experience a net drift equal to $\w2$ relative to the background plasma  \citep[see Figure 6 in][]{haggerty+20}. 
These drifts away from the shock lead to the removal of CR and magnetic energy from the shock and thus an enhancement of the fluid compression ratio and a steepening of the CR spectrum, as discussed in Section 5 of \cite{haggerty+20}. 

Equivalently, one can think of the postcursor as modifying the compression ratio ``felt" by CRs, just as the precursor modifies this ratio in \cite{caprioli12}. In the postcursor paradigm (ignoring, for now, the presence of a precursor), we have,

\begin{equation}
   \tilde{R} = \frac{u_1}{u_2+\w2} = \frac{R}{1+\alpha},
\end{equation}
where $\alpha \equiv \w2/u_2$. Thus, $q_{\rm p}$ depends only on $R$ and $\alpha$ or, equivalently, on $R$ and the magnetic pressure fraction downstream, $\xi_{\rm B,2} \equiv B_2^2/(8\pi\rho_0\vsh^2)$:
\begin{equation}
    q_{\rm p} = \frac{3R}{R-1-\alpha} = \frac{3R}{R-1-\sqrt{2R\xi_{\rm B,2}}}.
\end{equation}
Note that the effect of the postcursor will dominate that of a precursor, since compression of the magnetic field in the downstream leads to $\alpha > \w1/u_1$ \citep{caprioli+20}. 
In the case of efficient CR acceleration and thus magnetic field amplification, \cite{haggerty+20} reports $\alpha \sim 0.6$, which is sufficient to produce spectra steeper than $p^{-4}$, or $E^{-2}$ at relativistic energies. 
 
While these hybrid simulations provide a motivation and a physical explanation for the modification of the standard DSA theory, quantifying the steepening of the CR spectra in astrophysical systems requires additional calculations. 
Namely, the postcursor paradigm implies that spectral steepening increases with the downstream magnetic field strength, which, due to magnetic field amplification via CR-driven instabilities, increases with the CR pressure \citep[e.g., ][]{bell04,cristofari+21}. 
However, if spectra become too steep, the CR pressure will drop, reducing magnetic field amplification and thus causing the steepening to saturate.

In this paper, we use a semi-analytic model of NLDSA to generalize the results of \cite{caprioli+20} and estimate $q$ for a wide range of SNR shocks. 
We describe this model in detail in Section \ref{sec:method}. In Sections \ref{sec:results} and \ref{sec:discussion}, we present the results of our calculations and find that our modeled spectra produce good agreement with observations of both Galactic SNRs and radio SNe. We summarize in Section \ref{sec:conclusion}.

\section{Method} \label{sec:method}

To fully understand how a postcursor affects CR acceleration, we use a semi-analytic formalism to model SNR shocks over a range of ambient number densities, $\nism$, ambient magnetic fields, $B_0$, and SN energies, $E_{\rm SN}$. 
Herein we describe this formalism briefly, including our models for SNR evolution, particle acceleration, and magnetic field amplification. A more detailed description of our model, particularly our prescription for particle acceleration, can be found in \cite{caprioli12} and \cite{diesing+19}.

\subsection{Shock Hydrodynamics} \label{subsec:hydro}

We model SNR shock hydrodynamics using the formalism described in \cite{diesing+18}, which includes the effect of CR pressure on the evolution of the shock. 
More specifically, SNR evolution is modeled through three stages spanning $\gtrsim 10^5$ yr: the \emph{ejecta-dominated stage}, in which the mass of the swept-up ambient gas is less than that of the SN ejecta, the \emph{Sedov stage}, in which the swept-up mass dominates the total mass and the SNR expands adiabatically, and the \emph{pressure-driven snowplow}, in which the remnant cools due to forbidden atomic transitions but continues to expand because its internal pressure exceeds the ambient pressure. 
After this point, the remnant enters the \emph{momentum-driven snowplow}, in which the internal pressure falls below the ambient pressure and expansion continues due to momentum conservation. 

While we model SNRs through the end of the pressure-driven snowplow, the majority of CRs are accelerated during the transition between the ejecta-dominated and Sedov stages. 
The DSA timescale for CRs of energy $E = E_{\rm max}$ is given by $\tau_{\rm DSA} \approx D/\vsh^2$ where $D$ is the diffusion coefficient and $\vsh$ is the shock speed. 
Assuming Bohm diffusion \citep{caprioli+14c}, $D(E) \propto r_{\rm L} \propto E/B_2$ where $r_{\rm L} $ is the Larmor radius and $B_2$ is the postshock magnetic field. This gives $E_{\rm max} \propto B_2\vsh^2 t$ and, since $\vsh$ is roughly constant during the ejecta dominated stage, $E_{\rm max}$ initially increases. 
After the transition to the Sedov stage, the shock slows down such that $\vsh \propto t^{-3/5}$, meaning that $E_{\rm max}$ decreases with time, i.e., $E_{\rm max} \propto B_2(t)t^{-1/5}$ \citep{cardillo+15,bell+13}. 
The spectra of middle-aged and old SNRs are therefore most sensitive to shock evolution during the early adiabatic stages.

All SNRs are assumed to eject $M_{\rm ej} = 1M_{\odot}$ (1 solar mass) with $E_{\rm SN} \in [10^{51}, 10^{52}] \ \rm erg $ into a uniform ambient medium of density $\nism \in [10^{-1}, 10^{5}] \ \rm cm^{-3}$ and magnetic field $B_0 \in [3, 3000] \ \mu \rm G$.

\subsection{Particle Acceleration}

We model CR acceleration using a semi-analytic model of NLDSA described in \cite{caprioli+09a,caprioli+10b, caprioli12, diesing+19} and references therein, in particular \cite{malkov97,malkov+00,blasi02,blasi04,amato+05, amato+06}. 
This model self-consistently solves the diffusion-advection equation for the transport of non-thermal particles in a quasi-parallel, non-relativistic shock, including the dynamical backreaction of accelerated particles and of CR-generated magnetic turbulence.

Particles above a threshold in momentum, $\pinj$, are injected into the acceleration process, with $\pinj \equiv \xiinj\mpr \vsh/(1+\rt^{-1})$, consistent with the parameterization described in \cite{caprioli+15}, since  $\vsh/(1+\rt^{-1})$ is simply the velocity of the upstream fluid in the downstream frame. 
In general, an increase in $\xiinj$ corresponds to a decrease in the fraction of particles crossing the shock that are injected into DSA. Here we neglect the dependence of injection on the shock inclination and set an effective value of $\xiinj = 3.8$, which yields $\xicr \equiv P_{\rm CR}/(\rho_0 \vsh^2) \approx 0.1$ for a prototypical SNR ($\nism = 1$ cm$^{-3}$, $B_{0} = 3 \mu$G, $E_{\rm SN} = 10^{51}$ erg, $M_{\rm ej} = 1 M_\odot$) after a few hundred years, consistent with SNR observations. Note that $P_{\rm CR}$ refers to the CR pressure.

To account for the effects of a precursor and postcursor, we introduce into the diffusion-advection equation $\tilde{u}(x) \equiv u(x)\pm v_{\rm A}(x)$ , the effective fluid velocity as felt by the non-thermal particles which are scattered by magnetic structures moving at $v_{\rm A}(x)$ relative to the thermal plasma. 
Note that these structures move \emph{against} the fluid in the upstream, but \emph{with} the fluid in the downstream \citep[][]{caprioli+20}. 
Throughout this work, we assume that the postcursor extends beyond the diffusion length of the highest energy particles, i.e., behind the shock, $\tilde{u}(x) = u_2+\w2$. 

The actual extent of the postcursor in astrophysical shocks is difficult to quantify, even if high-resolution X-ray observations of individual SNRs with Chandra suggest that the magnetic field remains amplified on a scale of $1-5\%$ of the SNR radius \citep[e.g.,][]{tran+15}.
Physically speaking, since the maximum CR energy $E_{\rm max}$ is controlled by the smallest between the upstream and the downstream diffusion length \citep[e.g.,][]{drury83,lagage+83a,blasi+07}, the post-shock region with high magnetic field must be at least as extended as the diffusion length of particles with $E_{\rm max}$.
It follows that the postcursor must be more extended than the diffusion length of any particle, thereby leading to a \emph{global} steepening of the CR spectrum.
Note that, when only the Alfv\'enic drift in the precursor is retained  \citep[\'a la][]{zirakashvili+08b,caprioli12}, a global steepening is only possible if escaping CRs drive magnetic field amplification on all scales, which is not guaranteed.

In practice, our formalism begins with an initial guess for the CR pressure, which is used to solve the equations for conservation of mass, momentum, and energy across a plane, nonrelativistic shock. The magnetic field pressure, $P_{\rm B}$, is then calculated using the prescription described in \ref{subsec:mfa}, and the resulting $u(x)$ and $P_{\rm B}$ are then used to solve the diffusion-advection equation, which can be integrated to find a new guess for $P_{\rm CR}$. In this manner, our formalism iteratively solves for the CR spectrum while self-consistently accounting for the dynamical effect of accelerated particles and the amplification of magnetic fields. 

Once the proton spectrum has been calculated at each timestep of SNR evolution, particle momenta are shifted and the instantaneous spectra are weighted to account for adiabatic losses \citep[see][for more details]{caprioli+10a,morlino+12, diesing+19}. These weighted contributions are then added together to obtain a cumulative spectrum.

\subsection{Magnetic Field Amplification} \label{subsec:mfa}

The propagation of energetic particles ahead of the shock is expected to excite streaming instabilities, \citep[]{bell78a,bell04,amato+09,bykov+13}, which drive magnetic field amplification and enhance CR diffusion \citep{caprioli+14b,caprioli+14c}. 
The result is magnetic field perturbations with magnitudes that can exceed that of the ordered background magnetic field. 
This magnetic field amplification has been observationally inferred from the X-ray emission of many young SNRs, which exhibit narrow X-ray rims due to synchrotron losses by relativistic electrons \citep[e.g., ][]{parizot+06, bamba+05, morlino+10, ressler+14}. 

We model magnetic field amplification by assuming contributions from both the resonant streaming instability \citep[e.g., ][]{kulsrud+68,zweibel79,skilling75a, skilling75b, skilling75c, bell78a, lagage+83a}, and the non-resonant hybrid instability \citep{bell04}. A detailed discussion of these instabilities and their saturation points can be found in \cite{cristofari+21}.

In the resonant instability, CRs excite Alfv\'en waves with a wavelength matching their gyroradius. 
The growth of this instability saturates when the strength of magnetic perturbations reaches the level of the ordered background field: $\delta B/B \sim 1$. More specifically, \cite{amato+06} derives this saturation level to be,

\begin{equation}
    P_{\rm B1,res} = \frac{P_{\rm CR,1}}{4 M_{\rm A, 0}},
\end{equation}
where $M_{\rm A} \equiv \vsh/v_{\rm A,0}$ is the Alfv\'enic Mach number.

\begin{figure}
    \centering
    \includegraphics[width=0.5\textwidth, clip=true,trim= 25 10 40 35]{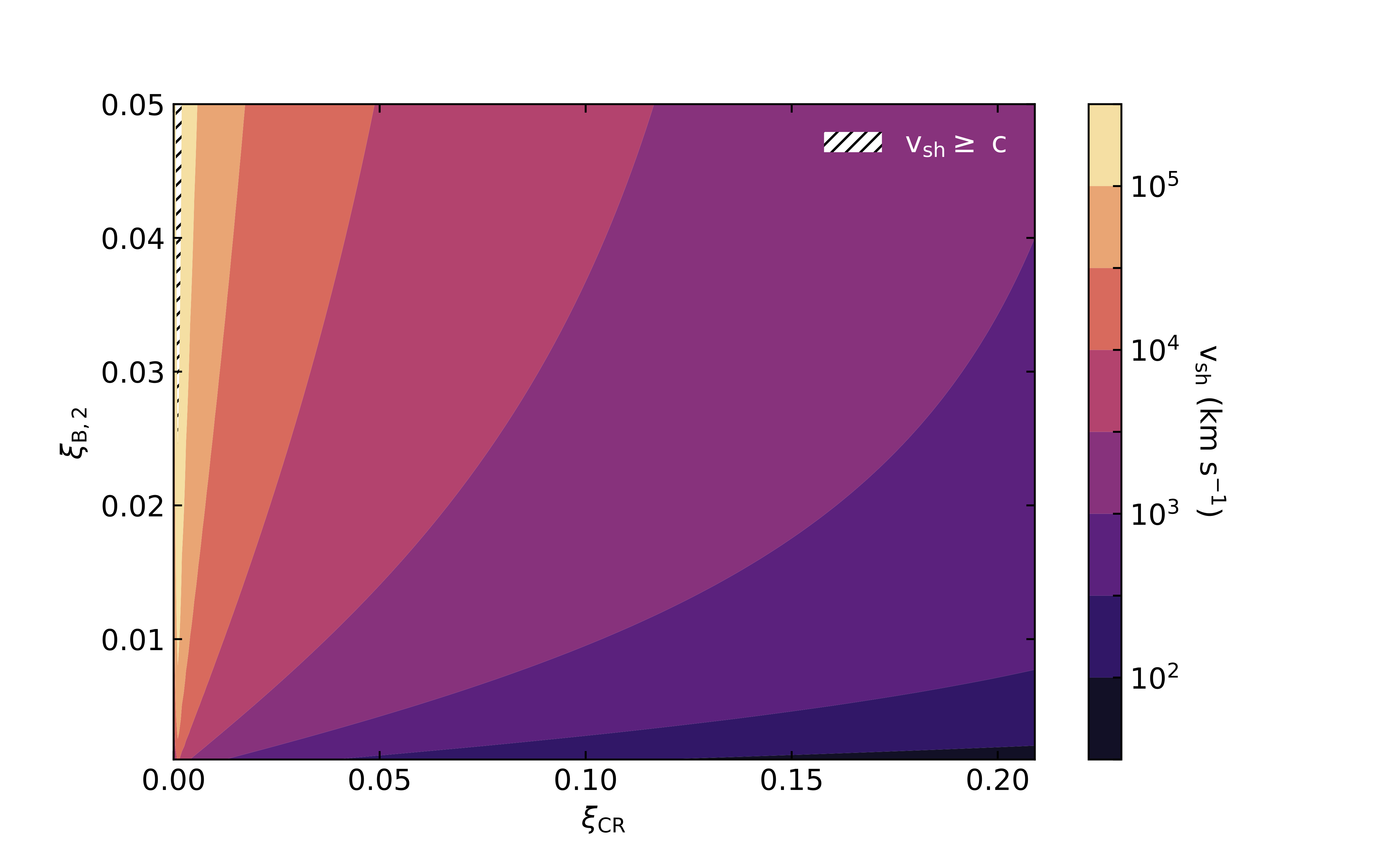}
    \caption{Shock velocity, denoted by color scale, as a function of CR acceleration efficiency, $\xicr$, and magnetic pressure fraction downstream, $\xi_{B,2}$, assuming a strong shock and magnetic field amplification dominated by the non-resonant streaming instability. Note that faster shocks correspond to stronger downstream magnetic fields given a fixed $\xicr$.}
    \label{fig:vsh_vs_xi}
\end{figure}

For fast shocks typical of young SNRs, more significant is the non-resonant hybrid instability. 
Driven by CR currents, $\bf{j}$, in the upstream,  \cite{bell04} predicts that saturation occurs when tension in magnetic field lines becomes sufficient to oppose the $\bf{j}\times\bf{B}$ force or, equivalently, when the magnetic field pressure reaches approximate equipartition with the anisotropic fraction of the CR pressure \citep[also see][]{blasi+15},

\begin{equation}
    P_{\rm B1,Bell} = \frac{\vsh}{2c}\frac{P_{\rm CR,1}}{\gamma_{\rm CR} - 1}.
\end{equation}
Here, $c$ is the speed of light and $\gamma_{\rm CR}=4/3$ is the CR adiabatic index. This saturation can lead to $\delta B/B_0 \gg 1$ and has been validated with hybrid simulations in \cite{zacharegkas+19p}. Thus, if the non-resonant instability dominates magnetic field amplification, one can solve the equations for conservation of mass, momentum, and energy across the shock to obtain a compression ratio (see Appendix \ref{sec:jumpcond}), yielding a well-defined relationship between the shock velocity, the CR acceleration efficiency ($\xicr$), and the magnetic pressure faction downstream, ($\xi_{\rm B,2}$). This relationship is shown in Figure \ref{fig:vsh_vs_xi} for a reasonable range of $\xicr$ and $\xi_{\rm B,2}$.

To account for both the resonant and non-resonant instabilities, we pose here that the upstream magnetic field pressure is given by $P_{\rm B,1} = \sqrt{P_{\rm B1,res}^2+P_{\rm B1,Bell}^2}$. 
The non-resonant instability dominates provided that $\vsh > \vsh^*$, where $\vsh^*$ is derived from the condition that $P_{\rm B1,res} = P_{\rm B1,Bell}$ and is given by,

\begin{equation}\label{eq:bell}
    \vsh^* = \sqrt{\frac{\w0 c}{6}}\simeq 572 \text{ km s}^{-1} \bigg(\frac{B_0}{3\mu \text{G}}\bigg)^{1/2}\bigg(\frac{n_0}{\text{cm}^{-3}}\bigg)^{-1/4}.
\end{equation}
$\vsh^*(n_0, B_0)$ is denoted with vertical lines in Figure \ref{fig:q_v}.
Assuming that all components of the magnetic perturbations upstream are compressed, the downstream magnetic field strength is $B_2 \simeq R_{\rm sub}B_1$. 

\begin{figure}
    \centering
    \includegraphics[width=0.5\textwidth, clip=true,trim= 25 10 40 35]{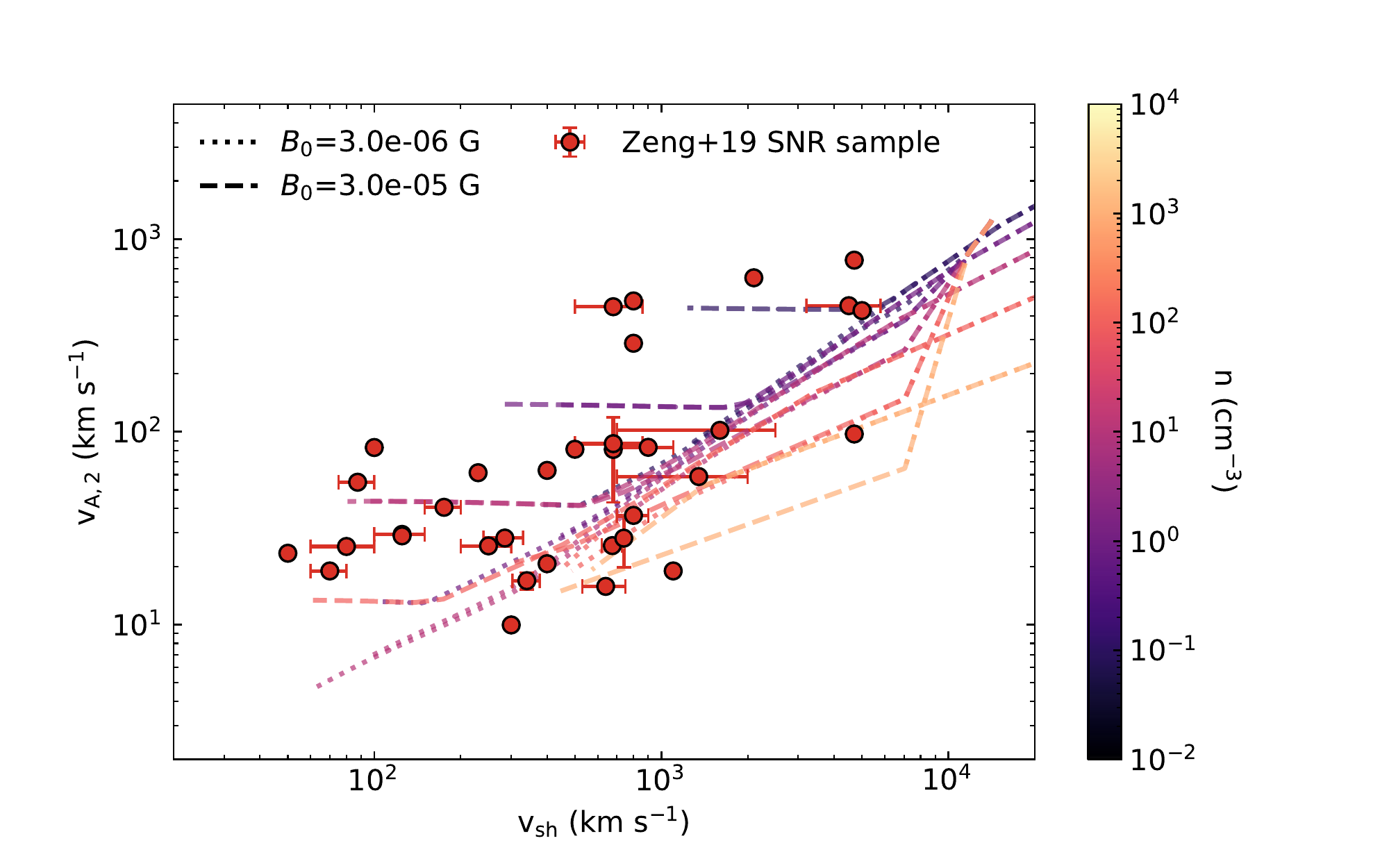}
    \caption{Downstream Alfv\'en speed, $\w2$, as a function of shock velocity, $\vsh$ for a number of modeled SNR evolutions (dotted and dashed lines). Each line corresponds to a single evolution with a fixed ambient density (color scale) and ambient magnetic field (line style). Overlaid are the SNR data aggregated in \cite{zeng+19}. Our prescription for magnetic field amplification produces modeled SNRs in good agreement with the measured relationship between $\w2$ and $\vsh$.}
    \label{fig:vA_vsh}
\end{figure}

Note that a comprehensive theory for magnetic field amplification upstream of a shock is still missing.
In particular, the relative contribution of escaping CRs \citep[e.g.,][]{vladimirov+06,caprioli+09b,bell+13} and diffusing CRs \citep[e.g.,][]{bell04,amato+06} may depend on their spectral slope.
The actual value of the field in the postcursor should, in principle, depend in a nonlinear way on the steepening that it induces \citep[see][for an extended discussion of these effects]{cristofari+21}. 
Such a self-regulating backreaction is not accounted for in the present calculation, but we check a posteriori that our prescription for magnetic field amplification is consistent with observations.

For an acceleration efficiency $\xicr \approx 0.1$, our typical SNR parameters give $B_2$ near a few hundred $\mu$G, in good agreement with X-ray observations of young SNRs \citep{volk+05,parizot+06,caprioli+08}. 

For a more robust test of our prescription, we consider the relationship between $\vsh$ and $\w2$. Specifically, our prescription predicts a positive relationship between $\vsh$ and $\w2$ for large $\vsh$ (i.e., where the non-resonant instability dominates). This relationship is independent of the ambient density and, for a strong shock with a weak precursor (i.e., $\rs = \rt = 4$), reads,

\begin{equation}
    \w2\simeq40 \text{ km s}^{-1}\bigg(\frac{\vsh}{1000 \text{ km s}^{-1}}\bigg)^{3/2}\bigg(\frac{\xicr}{0.1}\bigg)^{1/2}.
    \label{eq:va2}
\end{equation}
At lower $\vsh$ (i.e., where the resonant instability dominates), we would expect little to no correlation, since the resonant instability has a weaker dependence on $\vsh$ and depends on the ambient magnetic field, which may vary. In Figure \ref{fig:vA_vsh}, we compare our predicted relationship between $\w2$ and $\vsh$ to observational results compiled in \cite{zeng+19}. As Figure \ref{fig:vA_vsh} shows, our prescription yields a good agreement with observations. This agreement also provides circumstantial evidence that the presence of a postcursor is responsible for steep SNR spectra, particularly in light of the fact that SNRs with large $\vsh$ tend to have larger $q$ \citep[e.g., ][]{bell+11}.

\section{Results} \label{sec:results}

Herein we present our modeled CR spectra and quantify the steepening resulting from the modified shock dynamics--namely, the presence of a postcursor--described in \cite{haggerty+20} and \cite{caprioli+20}. 
Throughout this and subsequent sections, we estimate power-law slopes as,

\begin{equation}
    q\equiv -\left< \frac{d\log{\Phi(E)}}{d\log{E}}\right>,
\end{equation}
where $\Phi(E) = dN(E)/dE$ is the cumulative proton spectrum and $q$ is averaged between $10-10^3$ GeV.

\subsection{SNR Spectra}

\begin{figure}[t]
    \centering
    \includegraphics[width=0.5\textwidth, clip=true,trim= 10 10 10 35]{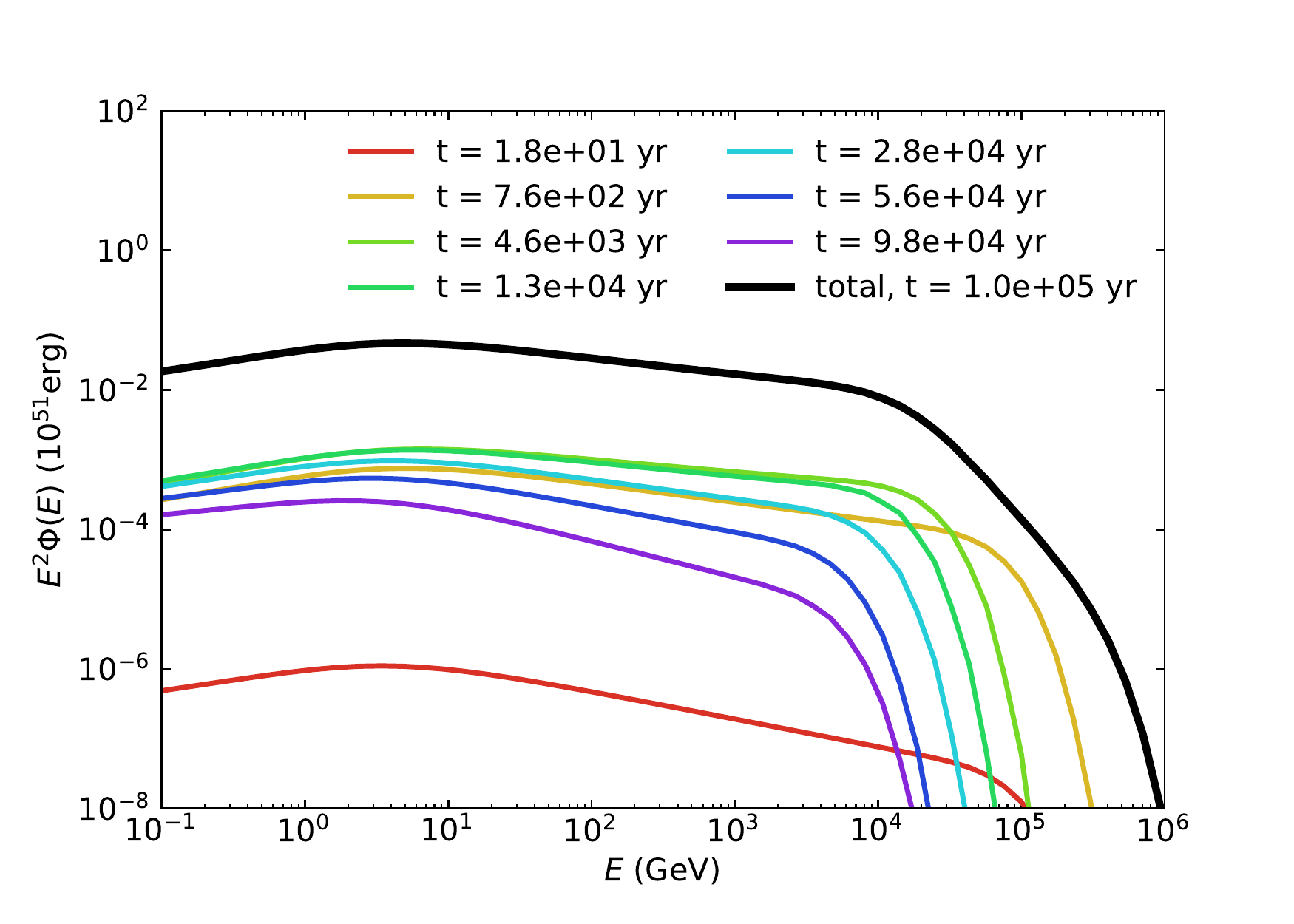}
    \caption{The modeled proton distribution, $\Phi(E)$, for a Tycho-like SNR: $\nism = 1$ cm$^{-3}$, $B_{0} = 3 \mu$G, $E_{\rm SN} = 10^{51}$ erg, and $M_{\rm ej} = 1 M_\odot$. The black line shows the cumulative proton spectrum after $10^5$ yr, while the colored lines show the contributions to this final spectrum from various timesteps. Throughout the SNR's evolution, protons are accelerated with spectra steeper than $E^{-2}$.}
    \label{fig:sample_spec}
\end{figure}

\begin{figure}[ht]
    \centering
    \includegraphics[width=0.5\textwidth, clip=true,trim= 10 10 10 35]{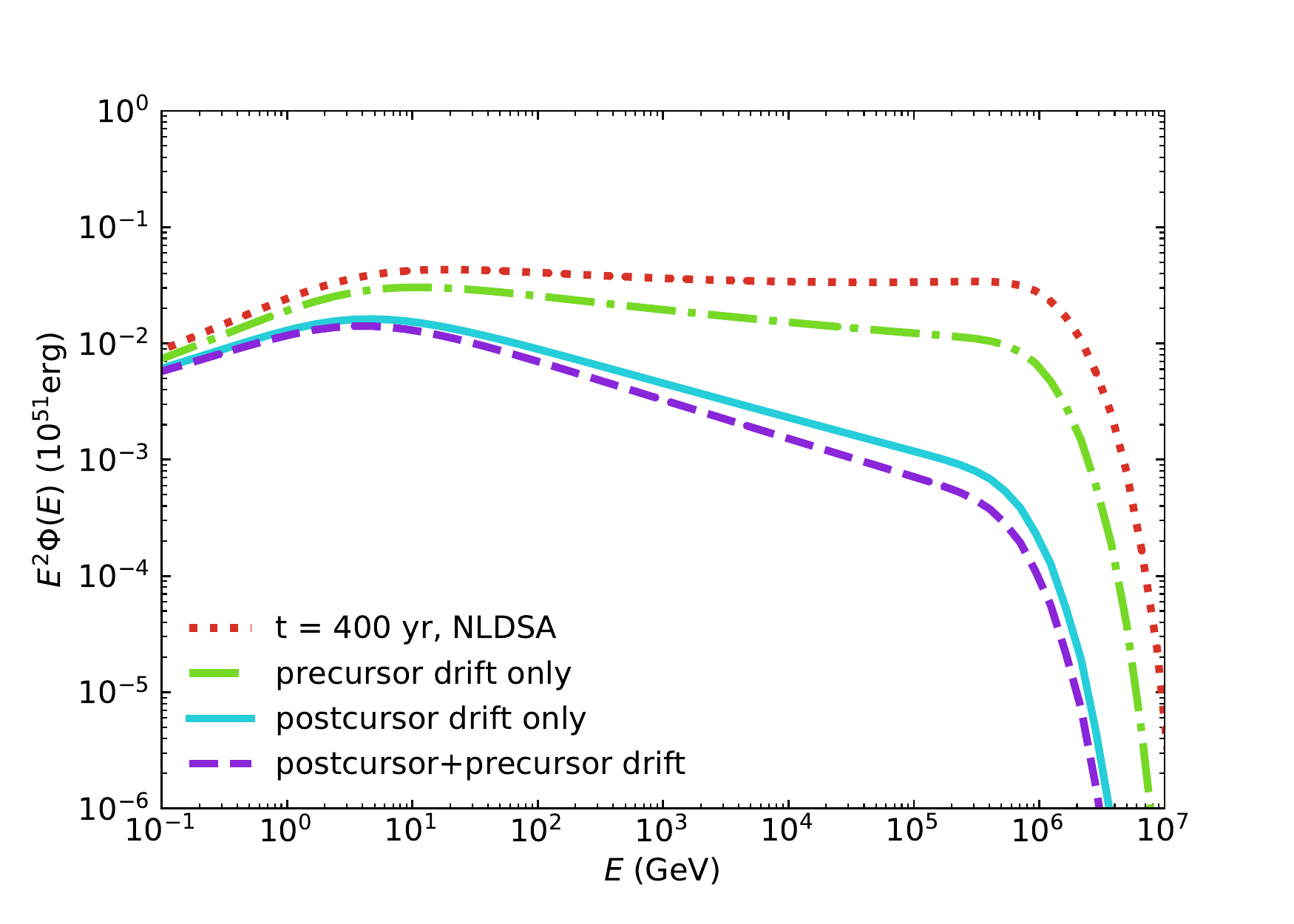}
    \caption{The modeled proton distribution of the Tycho-like SNR described in Figure \ref{fig:sample_spec} after 400 yr.
    Spectra are shown assuming traditional NLDSA with no net drift of magnetic fluctuations (red dotted line), assuming net drift in the precursor only (green dot-dashed line), assuming net drift in the postcursor only (blue solid line), and assuming the net drift in both the precursor and postcursor (purple dashed line). The inclusion of postcursor drift produces a substantial spectral steepening relative to the traditional NLDSA prediction. The addition of precursor drift further steepens the proton spectrum, but its effect is subdominant.}
    \label{fig:cursors}
\end{figure}

Our modeled spectrum of a ``prototypical,'' or Tycho-like SNR ($\nism = 1$ cm$^{-3}$, $B_{0} = 3 \mu$G, $E_{\rm SN} = 10^{51}$ erg, and $M_{\rm ej} = 1 M_\odot$) is shown in Figure \ref{fig:sample_spec}, including the contributions of protons accelerated at various stages of its evolution. These contributions are all steeper than $E^{-2}$, resulting in a cumulative spectrum $\Phi(E) \propto E^{-2.23}$ by the end of the SNR lifetime ($\sim 10^5$ yr). 

It is also worth noting that the slopes of these contributions do not vary monotonically. For the first $\sim 10^4$ yr, steepening due to the postcursor becomes less pronounced as the shock decelerates and the downstream magnetic field decreases (recall that $P_{\rm B1,Bell} \propto \vsh P_{\rm CR} \propto \vsh^3$, assuming the acceleration efficiency, $\xicr$ remains constant). After this point, $\vsh$ approaches a few hundred km s$^{-1}$ and the resonant streaming instability becomes the dominant source of magnetic field amplification. This instability has a weaker dependence on $\vsh$: $P_{\rm B1,res} \propto P_{\rm CR}/M_{\rm A,0} \propto \vsh$, again assuming constant $\xicr$. As a result, the spectrum stops hardening and actually begins to steepen slightly as $\xicr$ drops and the fluid compression ratio decreases. 
Note that, in our model, the drop in $\xicr$ is due to a decline of the postshock temperature and the ensuing decrease in the injection momentum ($\pinj \propto \vsh$), resulting in fewer of the GeV particles that are largely responsible for the CR pressure. 

\begin{figure*}[ht]
  \centering
  \subfloat[\label{fig:q_v_rhovar}]{%
      \includegraphics[width=0.5\textwidth, clip=true,trim= 35 10 40 35]{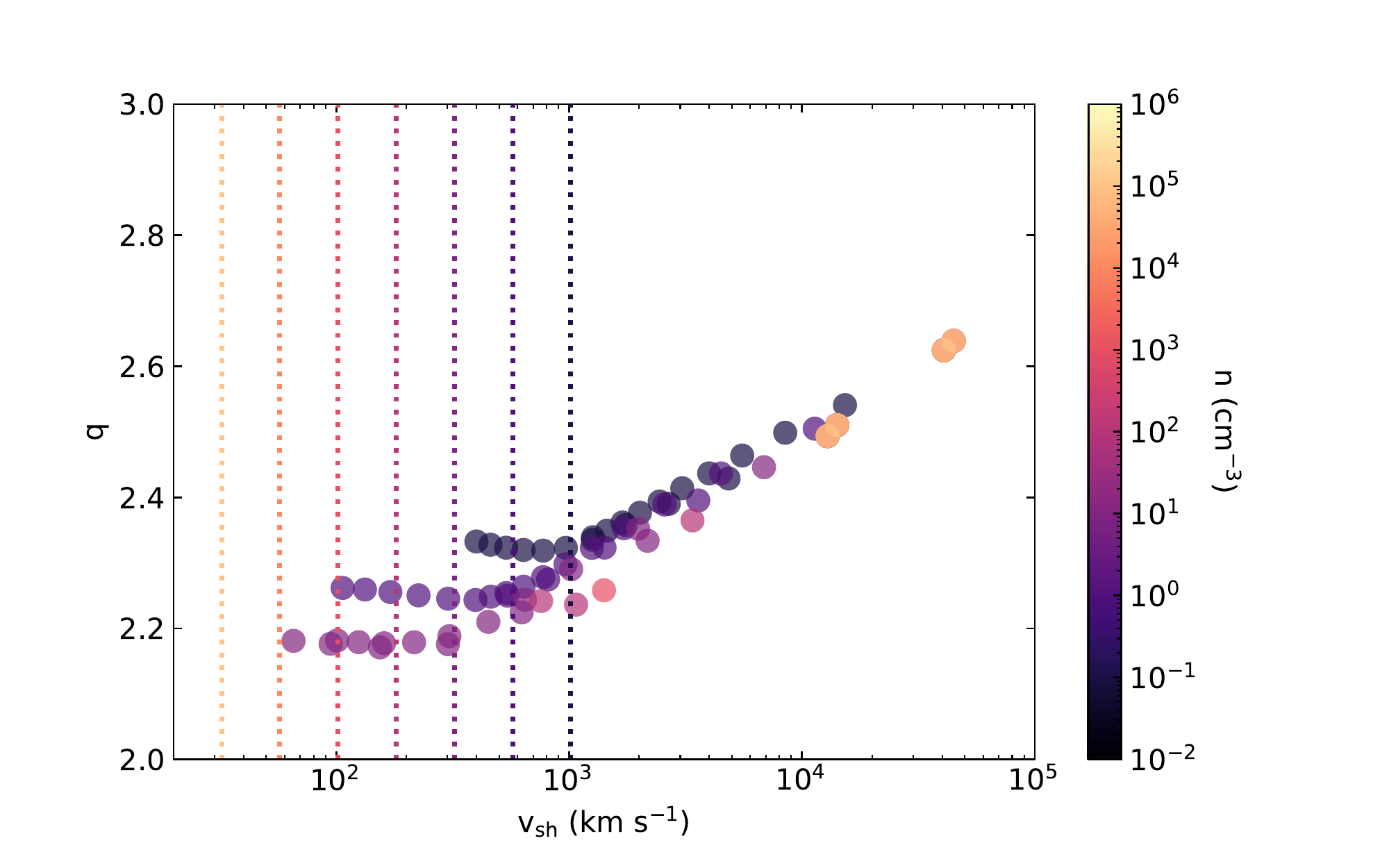}
    } 
    \subfloat[\label{fig:q_v_Bvar}]{%
      \includegraphics[width=0.5\textwidth, clip=true,trim= 35 10 40 35]{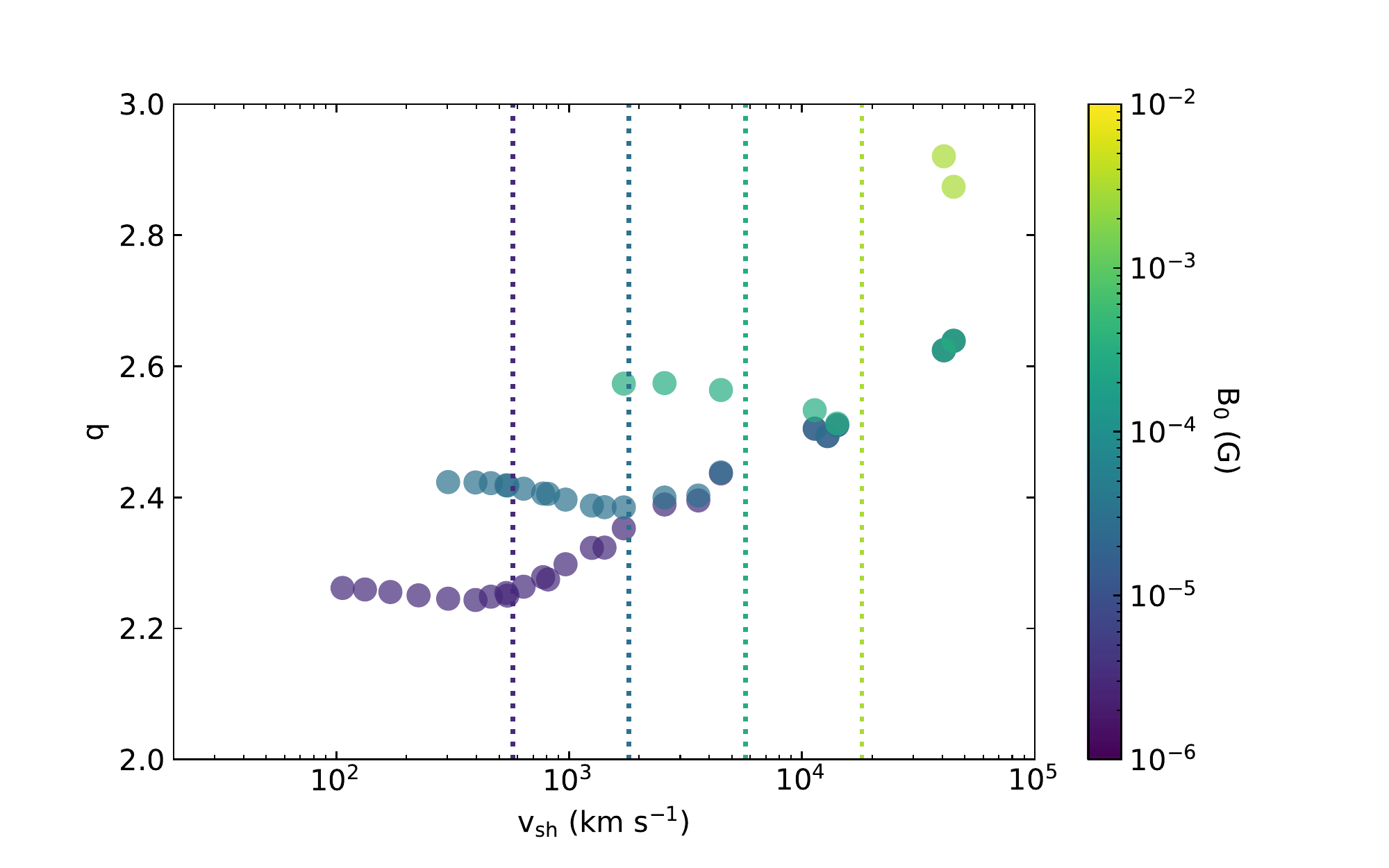}
    }

\caption{Power-law slopes, $q$, of modeled proton spectra as a function of shock velocity. The dotted vertical lines correspond to $\vsh^*$, the shock velocity where, for a given ambient density or magnetic field denoted by the color scale, the dominant source of magnetic field amplification transitions from the resonant to the non-resonant instability. \emph{Left}: The ambient magnetic field is held fixed at $3\mu$G while density, denoted by the color scale, is varied. \emph{Right}: The ambient density is held fixed at 1 cm$^{-3}$ while ambient magnetic field, again denoted by the color scale, is varied. In general, faster shocks give rise to larger magnetic field amplification and thus steeper spectra. However, this dependence on shock velocity disappears at low velocities where the resonant streaming instability is the primary source of magnetic field amplification.}
\label{fig:q_v}
\end{figure*}

The variation in the CR slope in conjunction with the decrease with time of $E_{\rm max}$ yields a cumulative proton distribution with a high-energy tail that is more extended than a simple exponential cutoff. Note that, with our prescription, this high-energy tail does not make it to PeV energies, or the ``knee" of the CR spectrum. 
This issue may be resolved by invoking a different class of SNR \citep[see, e.g., ][]{bell+13, cardillo+15, cristofari+21}.

\subsection{Spectral Steepening} \label{subsec:steepening}

A more explicit quantification of the effect of the postcursor can be found in Figure \ref{fig:cursors}. Here, we compare the cumulative spectrum of our Tycho-like SNR after 400 yr to the traditional NLDSA result and to the case with a postcursor but no net motion of magnetic structures in the precursor. As expected, the NLDSA formalism produces a modestly concave spectrum that deviates slightly from the standard $E^{-2}$ prediction. Meanwhile, the addition of a postcursor softens this spectrum substantially to $E^{-2.30}$. The addition of a precursor yields a slight increase in this steepening to $E^{-2.34}$, but its effect is underdominant due to the fact that the upstream magnetic field is decompressed such that $\w1/u_1 < \w2/u_2$. 

A summary of our results can be found in Figure \ref{fig:q_v}, which shows the average power law slope, $q$, as a function of shock velocity, $\vsh$, for the full range of modeled SNRs described in \ref{subsec:hydro}. To span a larger velocity range, we include models with initial energy, $E_{\rm SN}$, between $10^{51}$ and $10^{52}$ erg. Since increasing $E_{\rm SN}$ increases the shock velocity but does not otherwise affect shock hydrodynamics, we do not visually distinguish between different $E_{\rm SN}$ in Figure \ref{fig:q_v}. A fast shock may therefore correspond to a large $E_{\rm SN}$ or a young SNR; from the perspective of CR acceleration and magnetic field amplification, the two scenarios are equivalent. For this reason, our parameter range effectively spans different ejecta masses as well. Namely, an increase in $M_{\rm ej}$ simply corresponds to a decrease in $\vsh$ for a given $E_{\rm SN}$. With the range of SNR parameters described in \ref{subsec:hydro}, we obtain $2.1 \lesssim q \lesssim 3$.

For large $\vsh$ an increase in $\vsh$ corresponds to an increase in $q$, as one would expect when the Bell instability drives magnetic field amplification. 
As suggested in Figure \ref{fig:sample_spec}, this dependence disappears when $\vsh$ becomes small enough that the resonant instability dominates, i.e., at $\vsh^*$ (see Equation \ref{eq:bell}). Thus, the dependence of $\vsh^*$ on the ambient density and magnetic field introduces a spread in the relationship between $q$ and $\vsh$ which, in the case of small ambient densities and large magnetic fields, can extend up to high $\vsh$ ($\gtrsim 10^4$ km s$^{-1}$). More specifically, while $\nism$ and $B_0$ have no significant bearing on $q$ for $\vsh > \vsh^*$, they do determine the velocity below which $q$ becomes roughly constant or, equivalently, the minimum value of $q$ for a given SNR. An increase in $\nism$ yields a modest decrease in this minimum, since $\vsh^* \propto \nism^{-1/4}$, while an increase in $B_0$ increases this minimum, since $\vsh^* \propto B_0^{1/2}$. 

\section{Discussion}\label{sec:discussion}

\begin{figure}[t]
    \centering
    \includegraphics[width=0.5\textwidth, clip=true,trim= 25 10 40 35]{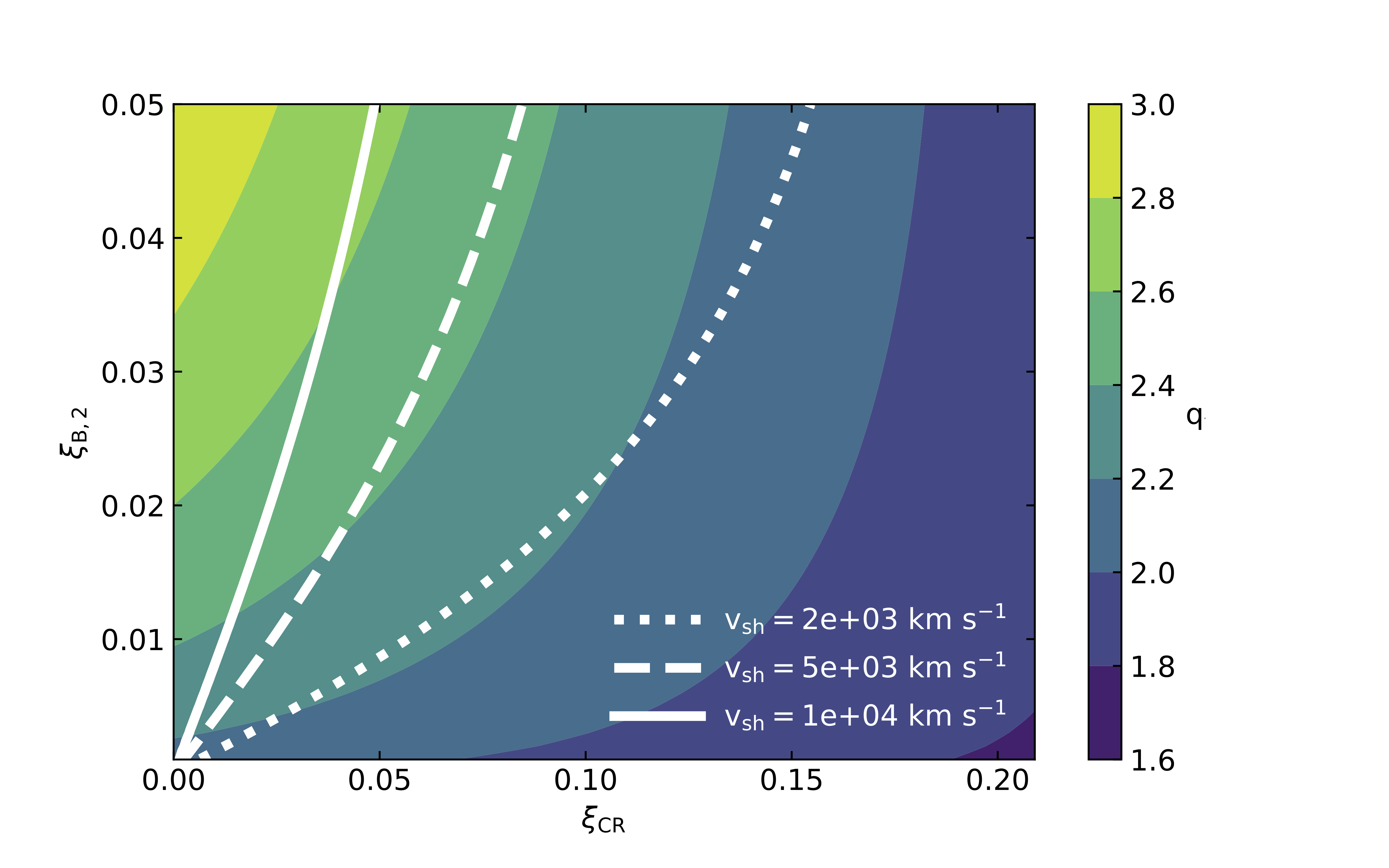}
    \caption{Predicted power law slope, $q$, denoted by color scale, as a function of CR acceleration efficiency, $\xicr$, and magnetic pressure fraction downstream, $\xi_{\rm B,2}$. $q$ is calculated for a strong shock assuming CRs probe the full compression ratio from the far upstream to the downstream (regions 0 and 2 respectively; see Figure \ref{fig:shock}). White lines denoting $\xi_{\rm B,2}$ as a function of $\xicr$ for various shock velocities are overlaid, assuming magnetic field amplification is dominated by the non-resonant instability. As explored in Section \ref{sec:results}, faster shocks correspond to steeper spectra for fixed $\xicr$.}
    \label{fig:q_vs_xi}
\end{figure}

By solving the equations for conservation of mass, momentum, and energy across a postcursor-modified shock, one can predict the fluid compression ratio as a function of the CR acceleration efficiency, $\xicr$, and the magnetic pressure fraction, $\xib$ (see Appendix \ref{sec:jumpcond} for details). 
Thus, the postcursor paradigm predicts a well-defined relationship between $q$, $\xicr$, and $\xib$. 
Assuming magnetic field amplification is driven by the non-resonant instability, $\xib$ can be recast in terms of $\xicr$ and $\vsh$, meaning that observational constraints on the shock velocity and spectral slope correspond to constraints on the CR acceleration efficiency.
For reference, we summarize this relationship in Figure \ref{fig:q_vs_xi}, assuming CRs probe the full compression ratio from the far upstream to the downstream, i.e., assuming CRs with energies above $\sim 1$ GeV. 
Less energetic CRs, such as those responsible for the synchrotron emission of radio SNe \citep{ellison+91, ellison+00, tatischeff09}, will probe smaller compression ratios and therefore exhibit slightly steeper spectra.

Equivalently, we can test the validity of the postcursor paradigm by comparing our predicted spectra to to observations, in particular the non-thermal emission of Galactic remnants (including historical SNRs), and young extragalactic supernovae (radio SNe). We find that the inclusion of a postcursor can reproduce both the modestly steep spectra of Galactic remnants ($\propto E^{-2.2}$) and the very steep spectra of radio SNe ($\propto E^{-3}$).

\subsection{Galactic Remnants}

\begin{figure}[t]
    \centering
    \includegraphics[width=0.5\textwidth, clip=true,trim= 35 10 40 35]{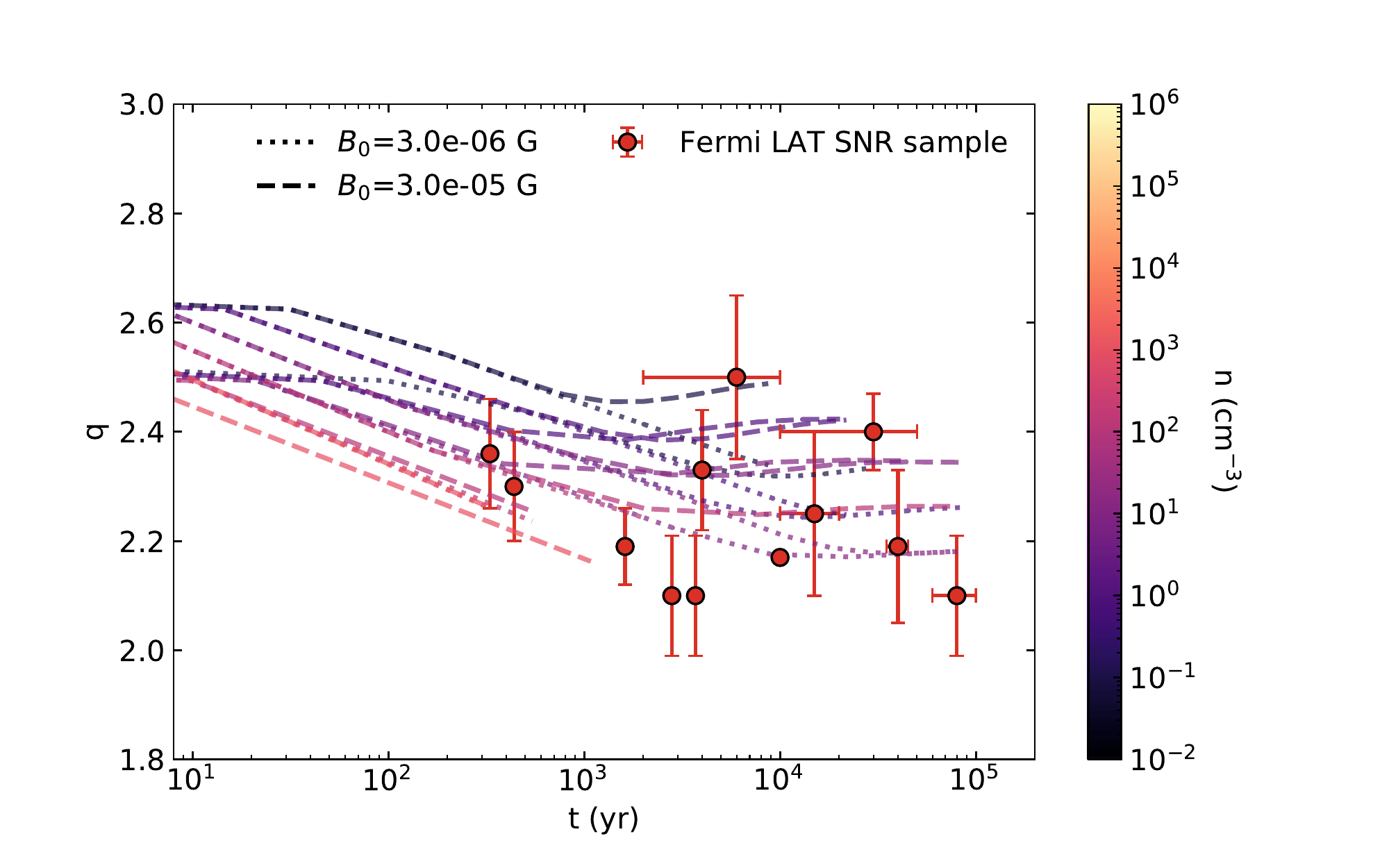}
    \caption{Power law slopes, $q$, of modeled proton spectra (dotted an dashed lines) as a function of SNR age. The slopes of GeV spectra from the Fermi LAT catalog \citep{acero+16short} are overlaid. For simplicity, SNRs have been removed if their GeV emission is likely leptonic in origin or exhibits a significant spectral break (see text for details). The spectral information for Cassiopeia A has been taken from \cite{saha+14}. The inclusion of a postcursor produces steep proton spectra in good agreement with SNR observations.}
    \label{fig:q_v_rhovar+data}
\end{figure}

The SNRs in our Galaxy consist largely of older, slower shocks \citep[$\vsh \ll 10^4$ km s$^{-1}$, see, e.g.,][]{green19}. 
Assuming magnetic field amplification driven by the non-resonant instability, we would therefore expect these SNRs--in the postcursor paradigm--to exhibit only modestly steep spectra. 

To test this, we look to GeV observations aggregated in \cite{caprioli11} from the Fermi LAT source catalog \citep{acero+16short}. 
We opt not to use synchrotron observations to avoid complications arising from cooling. Namely, synchrotron losses produce a steepening of the electron spectrum that depends strongly on the strength of the amplified magnetic field \citep[][]{diesing+19}. 
Meanwhile, TeV observations may probe the exponential cutoff of a CR distribution, artificially steepening the inferred slope. Admittedly, GeV observations suffer their own limitations. In particular, the emission process responsible for GeV photons may be $\pi_0$-decay (hadronic emission) or inverse-Compton (leptonic emission). However, for Galactic remnants, it is reasonable to assume that GeV emission with a spectral energy distribution (SED) steeper than $E^{-2}$ is hadronic in origin, since a leptonic origin would require an electron distribution steeper than $E^{-3}$ \cite[e.g., ][]{ghisellini13}. In principle, the bremsstrahlung radiation of relativistic electrons may also contribute to this GeV emission and lead to rather steep spectra, but it is usually underdominant with respect to hadronic emission for typical CR electron to proton ratios of $\lesssim 1\%$.
We therefore remove SNRs with spectra flatter than $E^{-2}$ for which a leptonic interpretation is favored: 
RX J1713.7-3946 \citep[e.g., ][]{HESS18} and Vela Jr. \citep[e.g., ][]{lee+13}. 
For simplicity, we also remove SNRs with breaks or cutoffs in the GeV range, which are typically interpreted as due to reacceleration \citep[e.g., W44, see][]{cardillo+16}. 
When possible, we use results from combined GeV-TeV analyses \cite[e.g., for Cas A,][]{saha+14}, which provide a more accurate representation of the full $\gamma$-ray slope.

Figure \ref{fig:q_v_rhovar+data} compares the values of $q$ calculated for our modeled SNRs to those in our sample. 
To simplify this figure, we do not include models with $B_0 > 30 \mu$G, since such strong fields are more typical of radio SNe (see Section \ref{subsec:RSN}) and are not required to reproduce these observations. 
Furthermore, $q$ is plotted against the estimated SNR age ($t$) rather than $\vsh$. 
As Figure \ref{fig:q_v_rhovar+data} shows, our models are able to reproduce the full range of slopes inferred from GeV and TeV observations of Galactic SNRs: $2.1 \lesssim q \lesssim 2.6$.

\subsection{Radio Supernovae} \label{subsec:RSN}

\begin{figure}[ht]
    \centering
    \includegraphics[width=0.5\textwidth, clip=true,trim= 10 10 10 35]{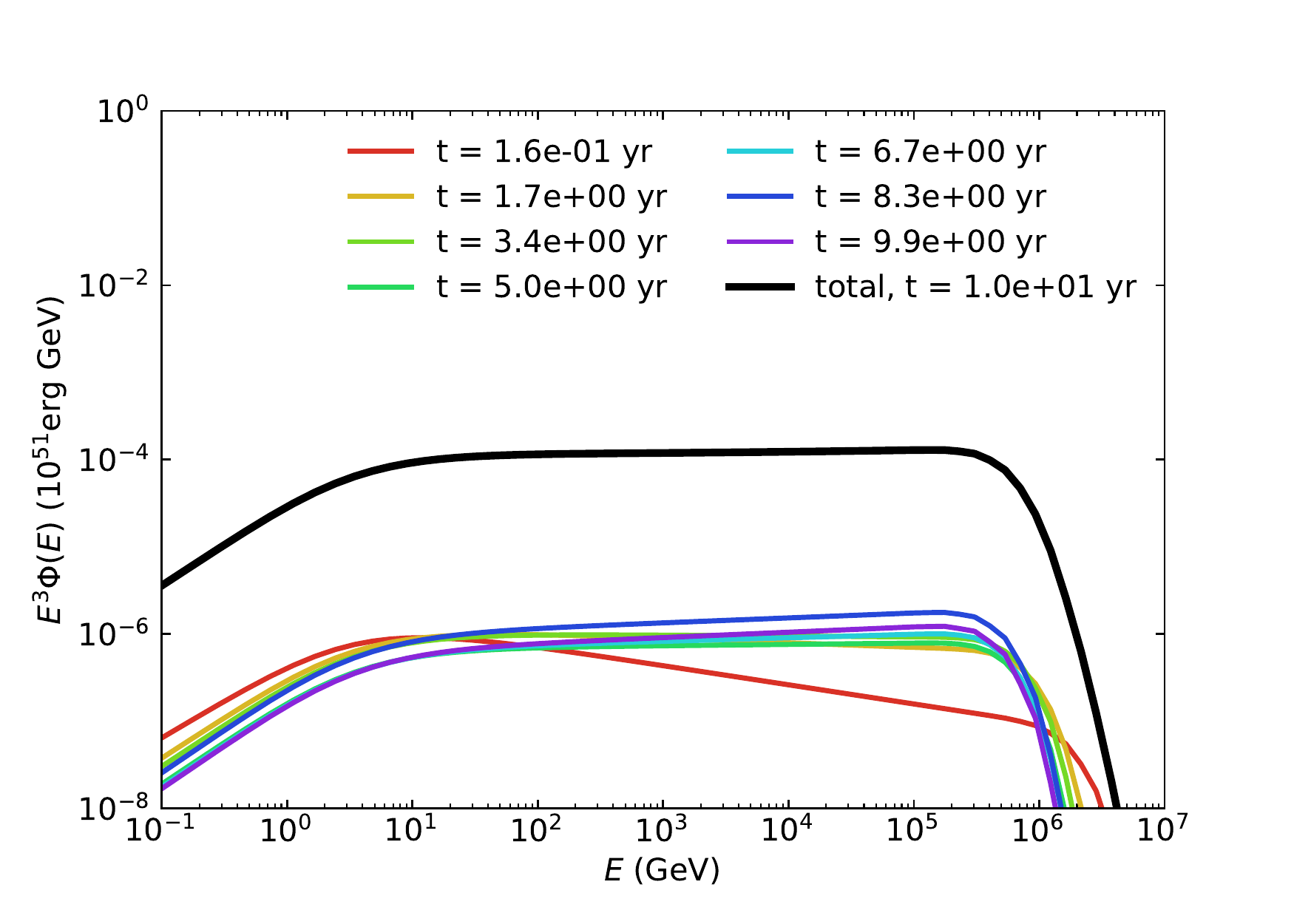}
    \caption{The modeled proton distribution ($E^3 \Phi(E)$) for a sample radio SN expanding into a circumstellar wind ($\nism \propto r^{-2}$; see text for details). The black line shows the cumulative proton spectrum after 10 yr, while the colored lines show the contributions to this spectrum from various timesteps. Our toy model reproduces the very steep spectra characteristic of radio SNe; for this setup, we obtain $q \simeq 2.99$.}
    \label{fig:sample_spec_RSN}
\end{figure}

In addition to explaining the modestly steep spectra of Galactic SNRs, the presence of a postcursor may also explain the very steep spectra of their extragalactic counterparts: radio SNe. 
These young, fast remnants ($\vsh \gtrsim 10^4$ km s$^{-1}$) typically expand into dense circumstellar winds blown by the progenitor star \citep[see, e.g., ][]{chevalier+17}. Their high $\vsh$ and, more explicitly, their large inferred postshock magnetic fields \citep[$\sim$ 0.1-1 G, see, e.g., ][]{chevalier98} imply strong magnetic field amplification, making them excellent candidates for tests of postcursor physics.

Intriguingly, radio SNe exhibit synchrotron emission that suggest electron distributions $\Phi(E) \propto E^{-3}$ or even steeper \citep[see, e.g., ][]{chevalier+06, soderberg+10, soderberg+12, kamble+16}. Assuming protons and electrons are accelerated with the same spectral slope--a reasonable assumption given that DSA depends only on a particle's rigidity--and that synchrotron cooling is negligible at energies corresponding to radio frequencies as discussed in \cite{chevalier+06}, we can conclude that the proton distribution must be similarly steep. 

As we have already discussed in Section \ref{subsec:steepening}, postcursor physics can reproduce $q \simeq 3$ under the right conditions: specifically, when $\vsh$ is large and the magnetic field is generated by the Bell instability. 
However, if our intent is to describe a typical radio SN, the models presented in \ref{subsec:steepening} are rather rough approximations, since they assume uniform ambient densities and include an injection prescription tuned to observations of Galactic SNRs (i.e., $\xiinj = 3.8$ so that $\xicr \approx 0.1$ for a prototypical Galactic remnant). To more accurately approximate the proton distribution of a typical radio SN, we produce a toy-model hydrodynamic evolution that follows an ejecta-dominated radio SN expanding into a circumstellar wind for approximately 10 years. We then use our semi-analytic formalism to self-consistently calculate the corresponding proton spectrum.

More explicitly, we consider an energetic SN ($E_{\rm SN} = 10^{52}$ erg) that ejects $M_{\rm ej} = 1 M_{\odot}$ into the circumstellar medium. Since we only model the first 10 years of evolution, the mass swept up by the shock is much smaller than $M_{\rm ej}$ and we therefore use the approximation in Table 9 of \cite{truelove+99} for an ejecta-dominated SNR expanding into a wind: $\vsh \propto t^{-1/5}$. 
For our circumstellar density, we assume a wind profile given by $\rho_0 = \dot{M}/(4\pi v_{\rm w})$, where $\dot{M}$ is the mass-loss rate of the progenitor and $v_{\rm w}$ is the wind velocity. As discussed in \cite{chevalier+06}, we assume typical paramenters for a Wolf-Rayet progenitor: $\dot{M} = 10^{-5} M_{\odot}$ yr$^{-1}$ and $v_{\rm w} = 1000$ km s$^{-1}$. We choose an ambient magnetic field that follows our density profile: $B_0/\text{G} \simeq 0.01 \sqrt{\nism/(5000 \text{ cm}^{-3})}$ with normalization chosen such that our magnetic field amplification prescription produces postshock fields consistent with observations \citep[$B_2 \sim$ 0.1-1 G, e.g.,  ][]{chevalier98}. Finally, since $\xiinj = 3.8$ gives extremely small acceleration efficiencies for our toy model ($\xicr < 0.01$), we reduce $\xiinj$ slightly to 3.4. With this adjustment, $\xicr$ remains modest ($ < 0.05$). The decrease in $\xiinj$ needed to produce acceleration efficiencies of 5-10\% would yield even steeper spectra.

Our model spectrum is shown in Figure \ref{fig:sample_spec_RSN} and has a slope of $q \simeq 2.99$;
note that to make this slope visually apparent, we plot $E^3\Phi(E)$.
As time passes, each new shell of protons contributes a slightly harder spectrum due to the modest decrease in $\vsh$, which leads to a reduction in magnetic field amplification (for the parameters discussed here, the non-resonant instability dominates). 
This behavior implies a simple physical explanation for the discrepancy between the very steep spectral slopes of radio SNe and the modestly steep slopes of Galactic SNRs. Namely, as young remnants age and slow down, their postshock magnetic fields decrease, reducing the strength of their postcursors and flattening their spectra. Of course, real radio SNe often exhibit spectra with more complex time variability, which may be attributed to circumstellar media that do not follow simple wind profiles \cite[e.g., ][]{margutti+19}. Our aim here is simply to show that, with reasonable parameters, the inclusion of a postcursor can easily reproduce the very steep spectra characteristic of radio SNe.

Our model predicts CR spectra to be steep even at high energies, while the classical concave-spectra explanation \citep[e.g., ][]{ellison+91, ellison+00, tatischeff09} returns rather flat spectra at TeV energies;
therefore, X-ray and, possibly, $\gamma$-ray observations may be able to distinguish between models.

\section{Conclusion} \label{sec:conclusion}
In summary, we use a semi-analytic model of NLDSA to quantify the CR spectral steepening in SNRs that arises from the presence of a \emph{postcursor}, i.e., a region behind a shock in which magnetic fluctuations drift away from the shock at the local Alfv\'en speed with respect to the background fluid. Since CRs isotropize with these fluctuations, they too experience a net drift, leading to a removal of CR energy from the system and thus a steepening of their spectra relative to the standard DSA prediction ($\Phi(E) \propto E^{-2}$). Our model also includes the effect of a \emph{precursor}, or region of enhanced CR density in front of the SNR shock. In this region, magnetic fluctuations move against the fluid (away from the shock) with the local Alfv\'en speed, leading to a further--albeit subdominant--steepening of the CR spectrum \citep[][]{caprioli12}. The formation of both a precursor and a postcursor has been validated with kinetic simulations \citep{haggerty+20, caprioli+20} and provides a natural explanation for the steep CR spectra inferred from observations of SNRs \citep[e.g., ][]{giordano+12, archambault+17, saha+14} and Galactic CRs, once corrected for propagation \cite[e.g.,][]{ams18,evoli+19a}.

Because magnetic fluctuations drift with the local Alfv\'en speed, it is important that we include a prescription for magnetic field amplification that is not only theoretically motivated, but consistent with observations. In our model, we implement a self-consistent prescription that incorporates the saturation points of both the resonant \citep[][]{amato+06} and non-resonant \citep[][]{bell04, zacharegkas+19p} streaming instabilities. This model yields magnetic fields that are consistent with those inferred from X-ray observations of young SNRs \citep[][]{vink+03,volk+05,parizot+06,caprioli+08} and reproduces the observed relationship between shock velocity and downstream Alfv\'en speed reported in \citet{zeng+19}.

With this prescription for magnetic field amplification \citep[also see][]{cristofari+21}, our model produces modestly steep spectra $\propto E^{-2.34}$ for a Tycho-like SNR after 400 yr: $\nism = 1$ cm$^{-3}$, $B_{0} = 3 \mu$G, $E_{\rm SN} = 10^{51}$ erg, and $M_{\rm ej} = 1 M_\odot$. 
We also confirm that the postcursor is the dominant source of this steepening; neglecting the effect of the precursor still yields spectra $\propto E^{-2.30}$.

As SNRs age and slow down, we find that this steepening diminishes, yielding a power-law slope, $q \simeq 2.23$ for our prototypical SNR after $10^5$ yr. Given observational constraints on the slope of the CR diffusion coefficient, this slope is consistent with that needed to reproduce the spectrum of Galactic CRs observed at Earth \cite[e.g., ][]{evoli+19a, evoli+19b}.

More generally, for large $\vsh$, the nonresonant instability dominates magnetic field amplification such that the magnetic pressure scales as $\vsh P_{\rm CR}$. As a result, the downstream Alfv\'en speed and thus the steepening due to the postcursor diminish as the SNR slows. This dependence largely disappears at lower $\vsh$, when the resonant instability dominates. The location of this transition, $\vsh^*$, depends on the ambient density and magnetic field (Equation \ref{eq:bell}).

The relationship between $\vsh$, magnetic field amplification, and $q$ that arises from postcursor physics provides a theoretically-motivated explanation for the modestly steep spectra of Galactic SNRs ($\propto E^{-2.2}$), the very steep spectra of radio SNe ($\propto E^{-3}$), and the connection between them. More specifically, we use our formalism to model both source classes and find that we are able to produce spectra in good agreement with observations. 

Our work represents the first generalization of postcursor physics to a wide range of SNR shocks, as well as the first self-consistent quantification of the spectral steepening that arises. The good agreement between our modeled spectra and those inferred from the nonthermal emission of real SNRs implies that the presence of a postcursor may resolve the tension between DSA predictions and observations. 

\acknowledgements
This research was partially supported by a Eugene and Niesje Parker Graduate Student Fellowship, NASA (grants NNX17AG30G and 80NSSC18K1726) and the NSF (grants AST-1909778, PHY-1748958, and PHY-2010240).

\begin{appendix}

\section{Solving the CR-Modified Jump Conditions}\label{sec:jumpcond}

We calculate the total compression ratio, $R_{\rm tot}$, by solving the equations for conservation of mass, momentum, and energy across the shock in a manner similar to that described in \cite{haggerty+20}. For a 1D, stationary shock, these equations read,

\begin{equation}
    \rho(x)u(x) = \rho_0u_0
\end{equation}
\begin{equation}
     \rho(x)u(x)^2 + P_{\rm g}(x) + P_{\rm B}(x) + P_{\rm CR}(x) = \rho_0u_0^2 + P_{\rm g,0}, \text{ and}
     \label{eq:pcons}
\end{equation}
\begin{equation}
    \frac{\rho(x)u^3(x)}{2} + F_{\rm g}(x) + F_{\rm B}(x) + F_{\rm CR}(x) =  \frac{\rho_0u_0^3}{2} + F_{\rm g,0} + F_{\rm CR,0},
    \label{eq:econs}
\end{equation}
where $F$ refers to the energy flux and subscripts g, B, and CR refer to the gas, magnetic field, and CR components respectively. We take the contribution of the magnetic field to be negligible in the far upstream, since the amplified field is typically much larger than the interstellar one.

However, while \cite{haggerty+20} closes this system of equations by neglecting the CR escape flux, $F_{\rm CR,0}$, we allow this flux to be nonzero. 
By using the canonical assumption that the CR distribution, $f(x,p)$, is continuous across the shock (i.e., that CRs have gyroradii large enough not to see the shock jump), all of the moments of $f(x,p)$ are also continuous;
this means that $P_{\rm CR}$ and $F_{\rm CR}$ are the same immediately upstream and downstream of the subshock. 
Thus, we solve Equation \ref{eq:econs} across the subshock rather than between downstream and upstream infinity, leaving $F_{\rm CR,0}$ to be defined by the solution of the CR transport equation.

The gas and magnetic energy fluxes can be written as in \cite{haggerty+20},
\begin{equation}
    F_{\rm g}(x) = \frac{\gamma_g}{\gamma_g-1}u(x)P_{\rm g}(x), \text{ and}
    \label{eq:geos}
\end{equation}
\begin{equation}
    F_{\rm B}(x) = (2\tilde{u}(x)+u(x))
    P_{\rm B}(x),
    \label{eq:Beos}
\end{equation}
where $\gamma_g = 5/3$ is the adiabatic index of the gas and $\tilde{u}(x)$ is the velocity of the magnetic fluctuations, $u(x) \pm v_{\rm A}(x)$. 
Note that Equation \ref{eq:Beos} assumes magnetic fluctuations are Alfv\'enic \citep{scholer+71, vainio+99, caprioli+09b};
while this is not strictly the case for turbulent fields, this prescription captures the fact that there are both Poynting and kinetic fluxes, and is empirically validated by kinetic simulations \citep[][]{haggerty+20}.
In the downstream, $\tilde{u}(x)$ becomes $u_2(1+\sqrt{2R_{\rm tot}\xi_{B,2}})$. For simplicity, we neglect the drift of magnetic fluctuations relative to the upstream flow, since $u_1 \gg \w1$.

Substituting Equations \ref{eq:geos} and \ref{eq:Beos} into Equation \ref{eq:econs}, dividing by $\rho_0u_0^3/2$, and assuming a gaseous subshock, we obtain,

\begin{equation}
    r^2+\frac{\eta_{\rm g}r^{-\gamma_{\rm g}}}{\gamma_{\rm g}M^2} + 6r\xi_{\rm B,1} = \frac{1}{R_{\rm tot}^2}+\frac{\eta_{\rm g}\xi_{\rm g,2}}{R_{\rm tot}}+(6+4\sqrt{2R_{\rm tot}\xi_{B,2}})\frac{\xi_{\rm B,2}}{R_{\rm tot}},
    \label{eq:econs_norm}
\end{equation}
where $r\equiv R_{\rm sub}/R_{\rm tot}$ and $\eta_g \equiv 2\gamma_{\rm g}/(\gamma_{\rm g}-1)$. Recall also that $\xi_i \equiv P_{\rm i}/(\rho_0 u_0^2)$. In the strong shock limit ($M \gg 1$), the second term on the left-hand side can be neglected.

Assuming a strong shock, $r$ and $\xi_{\rm g,2}$ can be rewritten in terms of known quantities by solving Equation \ref{eq:pcons}, normalized to $\rho_0u_0^2$, over the subshock and full shock, respectively:

\begin{equation}
    r = 1 - \xicr - \xi_{\rm B,1} \simeq 1 - \xicr, \text{ and}
    \label{eq:pcons_norm1}
\end{equation}
\begin{equation}
    \xi_{\rm g,2} = 1-\frac{1}{R_{\rm tot}}-\xicr-\xi_{\rm B,2}.
    \label{eq:pcons_norm2}
\end{equation}
We relate $\xi_{\rm B,1}$ and $\xi_{\rm B,2}$ by assuming that the magnetic field is compressed downstream: $\xi_{\rm B,2} = R_{\rm sub}^2\xi_{\rm B,1}$.

Equations \ref{eq:econs_norm}, \ref{eq:pcons_norm1}, and \ref{eq:pcons_norm2} can be combined into a single polynomial, written in terms of $x\equiv\sqrt{R_{\rm tot}}$:

\begin{equation}
    c_1x^4+c_2x^3+c_3x^2+c_4 = 0,
    \label{eq:poly}
\end{equation}
where,
\begin{equation}
\begin{split}
    &c_1 = -(1-\xicr)^2, \\
    &c_2 = 2^{5/2}\xi_{\rm B,2}^{3/2}, \\
    &c_3 = \eta_{\rm g}(1-\xicr-\xi_{\rm B,2})+6\xi_{\rm B,2}, \text{ and} \\
    &c_4 = 1-\eta_{\rm g}+\frac{6\xi_{\rm B,2}}{1-\xicr}.
\end{split}
\end{equation}
Here again we assume a strong shock for simplicity. We also write everything in terms of $\xi_{\rm B,2}$, in keeping with Figures \ref{fig:vsh_vs_xi} and \ref{fig:q_vs_xi}. It is also possible to solve for $R_{\rm tot}$ in terms of $\xi_{\rm B,1}$, but the resulting polynomial is substantially more complicated.

Equation \ref{eq:poly} has two positive roots, one of which corresponds to $1 \lesssim R_{\rm tot} \lesssim 2$;
the other corresponds to the physical solution used in the paper.

\end{appendix}


\bibliographystyle{aasjournal}

\end{document}